\documentclass[twocolumn,chaos,amsfonts,floatfix]{revtex4}
\ifx\pdfoutput\undefined
\usepackage[dvips]{graphicx}
\else
\usepackage[pdftex]{graphicx}

\usepackage[pdftex]{hyperref}
\fi
\graphicspath{{chaosrecfigures/LowRes/}}
\usepackage{amsmath,amsfonts,amssymb,amscd,amsthm,epsf}
\usepackage{subfigure}
\usepackage{color}
\newcommand{\enquote}[1]{``#1''}

\theoremstyle{remark}
\newtheorem{remark}{Remark} 

\newcommand{\ie}{{i.e., }}
\newcommand{\eg}{{e.g., }}
\newcommand{\Rset}{{\mathbb R}}
\newcommand{\Tset}{{\mathbb T}}
\newcommand{\nc}{\newcommand}
\nc{\figref}[1]{Fig.~\ref{fig:#1}}
\nc{\figsref}[2]{Figs.~\ref{fig:#1}-\ref{fig:#2}}
\nc{\tabref}[1]{Table~\ref{tab:#1}}
\nc{\tabsref}[2]{Tables~\ref{tab:#1}-\ref{tab:#2}}
\nc{\secref}[1]{Sec.~\ref{sec:#1}}
\nc{\secsref}[2]{Sec.~\ref{sec:#1}-Sec.~\ref{sec:#2}}
\nc{\ssecref}[1]{Sec.~\ref{ssec:#1}}
\nc{\ssecsref}[2]{Sec.~\ref{ssec:#1}-Sec.~\ref{ssec:#2}}
\nc{\eqeqref}[1]{Eq.~\eqref{eq:#1}}
\nc{\eqseqref}[2]{Eqs.~\eqref{eq:#1}-\eqref{eq:#2}}
\nc{\thmref}[1]{Theorem~\ref{theo:#1}}
\nc{\thmsref}[2]{Theorem~\ref{theo:#1}-\ref{theo:#2}}
\nc{\rcite}[1]{Ref.~\cite{#1}}
\nc{\rcites}[1]{Refs.~\cite{#1}}
\nc{\qtq}[1]{{\qquad\text{#1}\qquad}}
\nc{\vect}[1]{\boldsymbol{#1}}
\definecolor{changed}{rgb}{0.3,0.3,0.3}
\nc{\changed}[1]{{\bf previous version: }{\color {changed} #1}}

\begin{document}
\title{Meanders and Reconnection-Collision Sequences in the Standard
 Nontwist Map}
\author{A.~Wurm, A.~Apte, K.~Fuchss, and P.J.~Morrison}
\affiliation{Department of Physics and Institute for Fusion Studies\\
The University of Texas at Austin\\
Austin, TX 78712}
\date{\today}
\begin{abstract}
New global periodic orbit collision/separatrix reconnection scenarios
in the standard nontwist map in different regions of parameter space
are described in detail, including exact methods for determining 
reconnection thresholds that are implemented numerically. The results
are compared to a break-up diagram of shearless invariant curves.
 The existence of meanders (invariant tori that
are not graphs) is demonstrated numerically for {\it both} odd and
even period reconnection for certain regions in parameter space, and
some of the implications on transport are discussed.
\end{abstract}

\maketitle

{\bf  In recent years, area-preserving maps that violate the twist
condition locally in phase space have been the object of interest in
several studies in physics and mathematics. These {\it nontwist}
maps show up in a variety of physical models, e.g., in magnetic
field line models for reversed magnetic shear tokamaks. An important
problem is the determination and understanding of the transition to
global chaos (global transport) in these models. Nontwist maps
exhibit several different mechanisms: the break-up of invariant tori
and separatrix reconnections. The latter may or may not lead to
global transport depending on the region of parameter space. In this
paper we conduct a detailed study of newly discovered reconnection
scenarios in the {\it standard nontwist map}, investigating their
location in parameter space and their impact on global transport.}\\

\section{Introduction}
\label{sec:intro}

In this paper we consider the {\it standard nontwist map} (SNM) $M$,
as introduced in \rcite{del_castillo93}:
\begin{equation} \begin{split}
x_{n+1} & =  x_n + a \left(1-y^2_{n+1}\right)\,,\\
y_{n+1} & =  y_n - b \sin\left(2\pi x_n\right)\,,
\label{eq:stntmap}
\end{split} \end{equation}
where $(x,y) \in \Tset\times\Rset$ are phase space coordinates and
$a,b\in\Rset$ parameters. This map is {\it area-preserving} and
violates the {\it twist condition},
\begin{equation}
\frac{\partial x_{i+1}\left(x_i,y_i\right)}{\partial y_i}
 \neq 0\qquad \forall \ (x_i,y_i)\,,
\label{twist}
\end{equation}
along a curve in phase space, called the {\it nonmonotone
curve}.\cite{petrisor01} Some basic concepts used throughout the
paper are reviewed in Appendix~\ref{sec:basics}.

Nontwist maps are used to describe many physical systems, e.g.,
magnetic field lines in tokamaks (see, \eg \rcites{balescu98,
horton98, morrison00, oda95, petrisor03, stix76, ullmann00}) and
stellarators\cite{davidson95,hayashi95} (plasma physics); planetary
orbits,\cite{kyner68} stellar pulsations\cite{munteanu02}
(astronomy); traveling waves,\cite{weiss91,del_castillo93} coherent
structures and self-consistent transport\cite{del_castillo02} (fluid
dynamics). Additional references can be found in \rcites{apte03,
del_castillo96}. Apart from their physical importance, nontwist maps
are of mathematical interest because important theorems concerning
area-preserving maps assume the twist condition, \eg the KAM theorem
and Aubry-Mather theory. Up to now only few mathematical results are
known.\cite{apte04b,delshams00, franks03, moeckel90, simo98} Recently, it has
been shown\cite{dullin00, vanderweele88} that the twist condition is
violated generically in area-preserving maps that have a tripling
bifurcation of an elliptic fixed point. For studies of nontwist
Hamiltonian flows, we refer the interested reader to
\rcite{gaidashev04} and references therein.
\begin{remark}
  Although the SNM is not generic due to its symmetries (see
  Appendix~\ref{sec:symm}), it captures the essential features of
  nontwist systems with a local, approximately quadratic extremum of
  the winding number profile.
\end{remark}
Nontwist maps of the annulus exhibit interesting bifurcation
phenomena: periodic orbit collision and separatrix reconnection. The
former, which applies specifically to collisions of periodic orbits
of the same period, such as the so-called {\it up} and {\it down}
periodic orbits that occur in the SNM, can be used to calculate
torus destruction.\cite{del_castillo96} The latter is a global
bifurcation when the invariant manifolds of two or more distinct
hyperbolic orbits with the same rotation number connect leading to a
change in the phase space topology in the nontwist region. We
briefly review previous studies of reconnection in nontwist systems.

Howard and Hohs\cite{howard84} defined a quadratic nontwist map
closely related to \eqeqref{stntmap} and studied numerically the
reconnection of low-order resonances exhibiting homoclinic and
vortex-like structures.  Defining an average Hamiltonian they
predicted the reconnection threshold for period-one and period-two
fixed points. Howard and Humpherys extended the study to cubic and
quartic nontwist maps.\cite{howard95} These reconnection scenarios had
been conjectured by Stix\cite{stix76} in the context of the evolution
of magnetic surfaces in the nonlinear double-tearing instability, and
were seen by Gerasimov {\it et al.}\cite{gerasimov86} in a
two-dimensional model of the beam-beam interaction in a storage ring.

The first systematic study of reconnection was done by van der Weele
{\it et al.}\cite{vanderweele88,vanderweele90} in the context of
area-preserving maps with a quadratic extremum.  As far as we know the
terminology ``nontwist" and ``meanders" originates from there.

Del-Castillo-Negrete {\it et al.}\cite{del_castillo96} devised an
approximate criterion for the reconnection threshold of higher-order
resonances based on matching the slopes of the unstable manifolds of
the reconnecting hyperbolic orbits.

Continuing the work of Carvalho and Almeida,\cite{egydio92} Voyatzis
{\it et al.}\cite{voyatzis99} studied reconnection phenomena in
nontwist Hamiltonian systems under integrable perturbation with cubic
winding number profiles. Applying Melnikov's method, they showed the
transverse intersection of manifolds for arbitrarily small
nonintegrable perturbations for one reconnection scenario when the
winding number has a local extremum.\cite{voyatzis02}

The influence of manifold reconnection on diffusion was studied in the
region of strong chaos by Corso {\it et al.} in a series of
papers.\cite{corso98, corso99, corso03, prado00}

A very different criterion for the reconnection threshold was
proposed by Petrisor in \rcites{petrisor02,petrisor03}: If two
hyperbolic orbits have a heteroclinic connection, their {\it
actions} coincide. As noted in \rcite{petrisor02}, for odd-period
hyperbolic orbits in the SNM this criterion reduces to the action
being zero at the point of reconnection. This criterion was
implemented numerically in \rcite{wurm04} to estimate some
reconnection thresholds for odd-period orbits in the SNM.
\begin{remark}
  It has been noted that the above result about the equality of
  actions of reconnecting hyperbolic orbits is only approximately true
  in the near-integrable limit.\cite{llave04}
\end{remark}
For completion, we mention a few related studies: reconnection
phenomena and transition to chaos in the Harper map,\cite{saito97,
shinohara02} degenerate resonances in Hamiltonian systems with $3/2$
degrees of freedom,\cite{morozov02} and zero dispersion resonance in
the study of underdamped oscillators.\cite{soskin97}

Key to the analytical and numerical exploration of the standard
nontwist map is the map's invariance under symmetries, reviewed in
Appendix~\ref{sec:symm}. Of particular significance are the
\emph{indicator points},\cite{shinohara98} fixed points of some of
the symmetries of the SNM, whose importance was first recognized by
Shinohara and Aizawa. These points were independently re-discovered
by Petrisor\cite{petrisor01} in the analysis of the reversing
symmetry group\cite{lamb92} of nontwist standard-like
area-preserving maps. In \rcite{shinohara97}, it was shown that a
shearless invariant torus crosses the $x$-axis at two points. This
led the authors to devise a criterion to determine the approximate
location in the $(a,b)$-parameter space of the break-up of shearless
invariant tori for many winding numbers (see \ssecref{nt-breakup}).

Subsequently, based on numerical observations, Shinohara and
Aizawa\cite{shinohara98} used indicator points to propose exact
expressions for the collision threshold of even-period orbits and a
method to determine numerically the reconnection threshold for
odd-period hyperbolic orbits.

The goal of the present paper is to describe reconnection-collision phenomena
in detail in {\it all} regions of $(a,b)$-parameter space for the
standard nontwist map.  The details of the two main
scenarios (odd-period and even-period) depend crucially on the
$(a,b)$-region, and have not been discussed exhaustively up to now.
The main tool we use here is the numerical implementation of
two criteria for collision thresholds: the analytic one of
\rcite{shinohara98} and the numerical one of \rcite{del_castillo96}.
The resulting curves are compared with the $(a,b)$-space break-up
diagram (see \ssecref{nt-breakup}), produced by a modified version
of \rcites{apte03, shinohara97}. Early results of this investigation
were reported in \rcite{wurm04}, and some of them will be elaborated
below for completeness.

The paper is organized as follows. We review some basic concepts of
nontwist systems and the SNM in \secref{nontwist}.  Some novel
reconnection and collision scenarios for orbits of even and odd
periods are described in \secref{evenreco} and \secref{oddreco},
respectively, applying and extending methods from \rcites{del_castillo96,
 shinohara98}. The results are discussed in the context
of the break-up diagram in \secref{discussion}. In \secref{conclusion} we give our
conclusions and indicate some directions of future research. The appendices contain basic
definitions and a brief summary of symmetry properties of the SNM.

\section{Review of nontwist maps}
\label{sec:nontwist}

\begin{figure} \begin{center}
\includegraphics[width=3in]{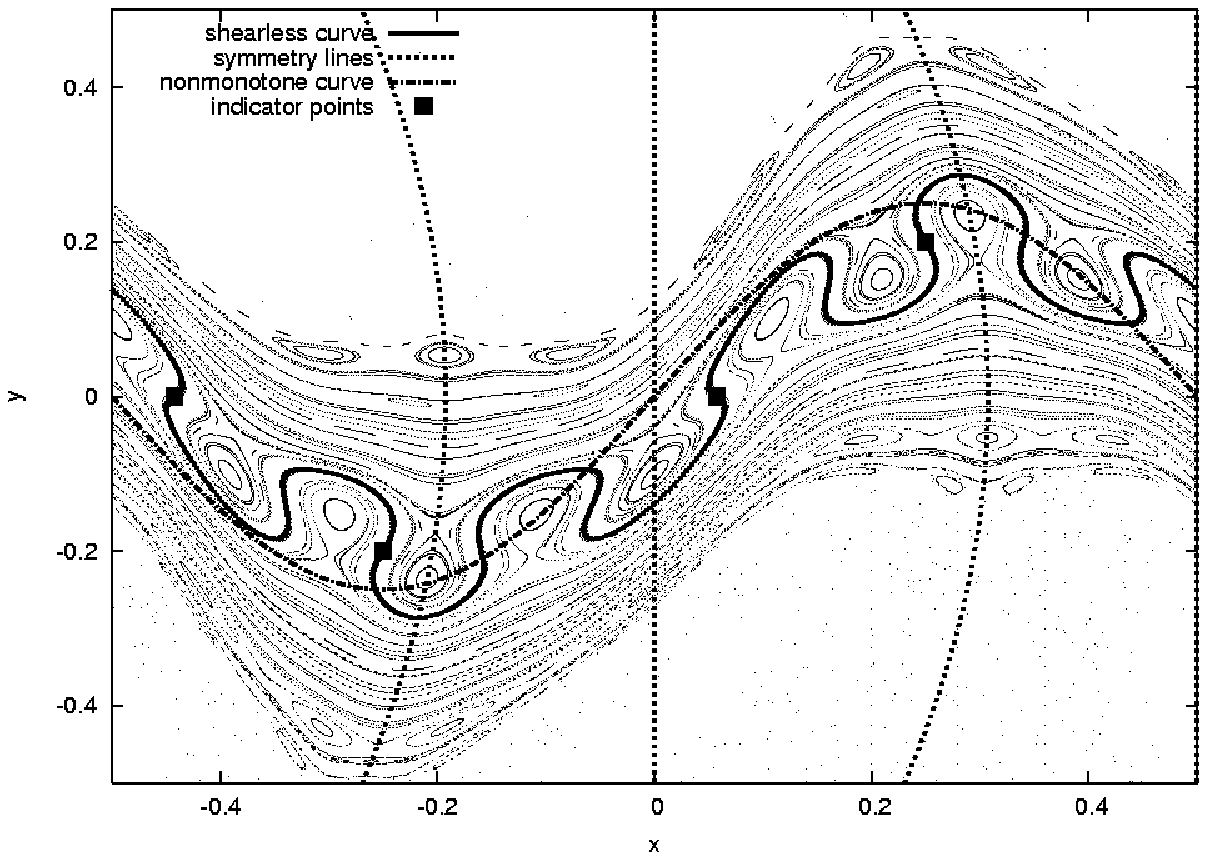}\\
\includegraphics[width=3in]{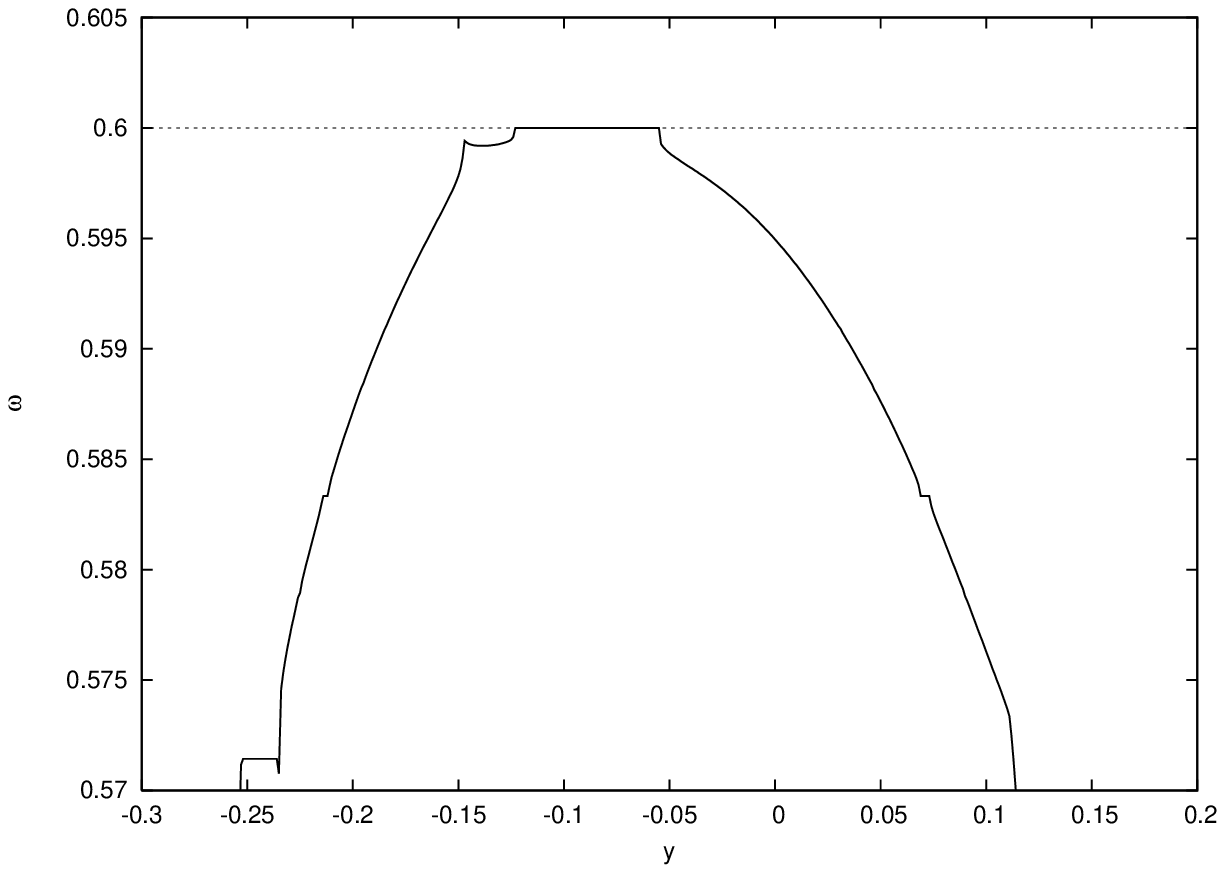}
\caption{Example of the phase space for standard nontwist map (top)
  and winding number profile along $y$-axis (bottom) at $a=0.615$,
  $b=0.4$. The symmetry lines, nonmonotone curve $\cal C$, $\cal
  G$-invariant curve $\gamma _S$, and indicator points are
  displayed.}
\label{fig:samplexy} \end{center} \end{figure}
In this section, we review some fundamental concepts that
are required for further explorations of the SNM. \figref{samplexy} shows a
typical phase space plot along with the plot of winding number $\omega$ versus
$y$ (henceforth called \emph{winding number profile}) for the central
section of the $y$-axis.

The phase space of the SNM consists of two twist regions, ``above''
and ``below'' the center, \ie regions in which Eq.~(\ref{twist}) is
satisfied. Since the twist condition is violated along the nonmonotone
curve $\cal C$ given by $y=b\sin(2\pi x)$, only orbits with points
falling on $\cal C$, referred to as {\it nonmonotone} orbits by
Petrisor,\cite{petrisor01} and orbits with points on both sides of
$\cal C$, called {\it pseudomonotone}, are affected by the nontwist
property. Note that, for any $b\neq 0$, the curve $\cal C$ is not an invariant
torus.

Like the phase space plot, the winding number profile in the outer
regions looks the same as that of a twist map, showing, \eg the
familiar plateaus associated with islands around periodic orbits
(here orbits with winding numbers 3/5, 7/12, and 4/7). The only
distinctly nontwist effect, aside from the existence of the overall
maximum (here at 3/5) and of multiple periodic orbits of winding
numbers less than the maximum, is the valley below the 3/5 plateau,
which gives rise to a variety of phenomena discussed in
\secref{nonstd}.

\subsection{Reversing symmetry group and shearless curve}
\label{ssec:nt-symm}

When the winding number profile has a local extremum at an irrational
value of $\omega$, the corresponding invariant torus is called
{\it shearless}. As discussed by Petrisor,\cite{petrisor01} the SNM has
 at most one homotopically nontrivial shearless invariant torus
 $\gamma_S$ that is also invariant under the reversing
symmetry group $\mathcal{G}$ (reviewed in Appendix~\ref{sec:symm}).
When it exists, $\gamma_S$ also contains the indicator points
\begin{equation}
  \vect{z}_0^{(0,1)} = \left(\mp\frac{1}{4},\mp\frac{b}{2}\right)\,,
  \mbox{ and } \vect{z}_1^{(0,1)} = \left(\frac{a}{2}\mp\frac{1}{4} ,
  0\right)\,, \label{eq:shin_pts}
\end{equation}
and all their iterates. As we will discuss in \secref{nonstd}, in some
regions of parameter space, there exist several shearless tori, but only
one of them is $\mathcal{G}$-invariant.

  The significance of a shearless torus is
that it acts as a barrier to transport, whereas the nonmonotone
curve $\cal C$ does not. \figref{samplexy} shows the nonmonotone curve and
the shearless torus, along with symmetry lines and indicator points.

The standard nontwist map is non-generic because it is time-reversal
symmetric as well as invariant under a symmetry, as reviewed in
Appendix~\ref{sec:symm}. Nevertheless, most of the phase space
phenomena described in this paper are also observed in arbitrary
nontwist systems, even though the exact definitions of many of the
concepts introduced to study them cannot be generalized to these systems.

\subsection{Periodic orbit collision and standard separatrix reconnection}
\label{ssec:nt-coll}

As mentioned earlier, one consequence of the violation of the twist
condition is that the SNM has more than one orbit (either invariant
tori or chains of periodic orbits) of the same winding number. These
orbits can collide and annihilate at certain parameter values.
\begin{figure*}[t] \begin{center}
\includegraphics[width=0.7\textwidth]{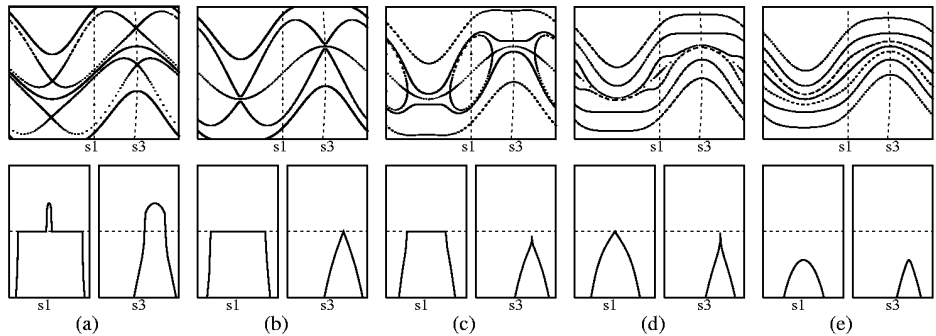}
\caption{The standard scenario of even-period
  reconnection/collision sequences (upper row, $y$ vs.\ $x$) and winding number profiles along
  the $s_1$ and $s_3$ symmetry lines (lower row, $\omega$ along $s_i$
 vs.\ $y$). (discussion in text)}
\label{fig:evenstd}
\end{center}
\end{figure*}
The collision of periodic orbits involves another purely nontwist
phenomenon, namely the reconnection of the invariant manifolds of the
corresponding hyperbolic orbits. These reconnection-collision
sequences in the SNM are distinctly different for orbits of even and
odd periods.  Here we will only give a brief account of the simplest
version of both sequences, which we will refer to as the {\em
standard scenarios}.\cite{vanderweele90} We show phase space plots
and winding number profiles for several steps in these sequences. The
 description of more intricate reconnection-collision scenarios
appears in Secs.~\ref{sec:evenreco} and~\ref{sec:oddreco}.
\begin{figure*}[t] \begin{center}
\includegraphics[width=0.7\textwidth]{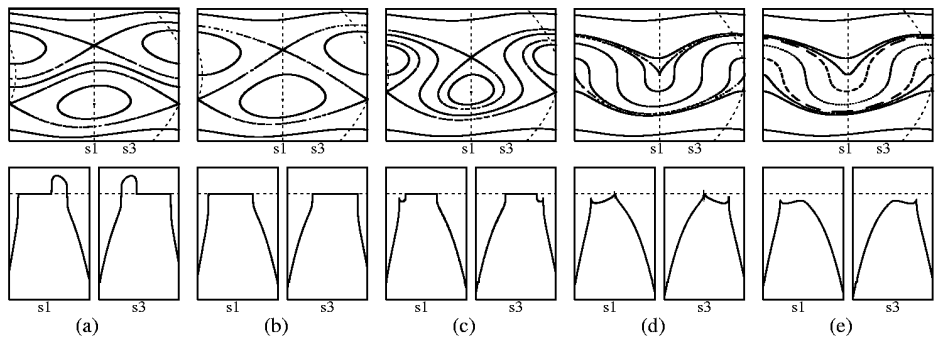}
\caption{The standard scenario of odd-period
  reconnection/collision sequences (upper row, $y$ vs.\ $x$) and winding number profiles along
  the $s_1$ and $s_3$ symmetry lines (lower row, $\omega$ along $s_i$
vs.\ $y$). (discussion in text)}
\label{fig:oddstd}
\end{center}
\end{figure*}
Changing the parameter values of the SNM we see the following sequences:
\begin{itemize}
\item For orbits with even period, i.e., with the same stability type for
  both the up and down orbits on a symmetry line
  (Fig.~\ref{fig:evenstd}a): collision of the hyperbolic orbits
  which is also the threshold for reconnection
  (Fig.~\ref{fig:evenstd}b); the ``dipole topology'' in which the
  hyperbolic orbits have moved off the symmetry lines
  (Fig.~\ref{fig:evenstd}c); the collision of elliptic orbits that
  coincides with the collision of the non-symmetric hyperbolic orbits
  (Fig.~\ref{fig:evenstd}d) leading to annihilation of these
  periodic orbits (Fig.~\ref{fig:evenstd}e).\\
  The winding number profile shows a maximum that is greater than,
  equal to, and less than the winding number of the periodic orbit
  before, during, and after this process, respectively.
\item For orbits with odd period, i.e., with opposite stability type for the
  up and down orbits on a symmetry line
  (Fig.~\ref{fig:oddstd}a): reconnection of hyperbolic manifolds
  of up and down orbits (Fig.~\ref{fig:oddstd}b); appearance of
  non-KAM \emph{meandering} orbits (elaborated on in
  \ssecref{nt-meanders}), homoclinic separatrices, and
  \emph{dimerized} chains (Fig.~\ref{fig:oddstd}c);
  hyperbolic-elliptic collision (Fig.~\ref{fig:oddstd}d) leading
  to annihilation of these periodic orbits
  (Fig.~\ref{fig:oddstd}e).\\
  As above, the winding number profile shows a global maximum that is
  greater than, equal to, and less than the winding number of the
  periodic orbit before, during, and after this process,
  respectively. But in addition, there is a local minimum -- associated
  with the appearance of the meandering orbits -- which persists even
  after the collision. (See~\ssecref{nt-meanders}.)
\end{itemize}

\subsection{Bifurcation and indicator curves}
\label{ssec:bif_curves}

The parameter values for the threshold of collision of periodic
orbits with a fixed winding number are numerically observed to lie
on a smooth curve in $(a,b)$-parameter space, called {\it bifurcation curve},
 which was first defined in \rcite{del_castillo96}.

The {\it $m/n$-bifurcation curve $b=\Phi_{m/n,i}(a)$} is the set of
$(a,b)$ values for which the $m/n$ up and down periodic orbits on the
symmetry line $s_i$ (see Appendix~\ref{sec:symm}) are at the point
of collision. (For clarity, the figures in this paper denote these
curves by ``bc1,'' ``bc2'' etc. for symmetry lines $s_1$, $s_2$
etc.) The main property of this curve is that for $(a,b)$ values
below $b=\Phi_{m/n,i}(a)$, the $r/s$ periodic orbits, with $r/s <
m/n$, exist. Thus, $m/n$ is the maximum winding number for parameter
values along the $m/n$-bifurcation curve.

For $n$ odd, bifurcation curves along $s_1$ through $s_4$ coincide
in the SNM because of the map's high degree of symmetry. For $n$
even, bifurcation curves along $s_1$ and $s_2$ are separate from the
ones along $s_3$ and $s_4$ for any $b\neq 0$.
\begin{table*}
\begin{center}
\begin{tabular*}{0.7\textwidth}{@{\extracolsep{\fill}}ccc}
\hline
 & Elliptic collision & Hyperbolic reconnection \\
\hline
$\omega=\frac{1}{2}$
    & $a=\frac{1}{2}$
    & $b^2 = 4\left(1-\frac{1}{2a}\right)$ \\
$\omega=\frac{1}{4}$
    & $b^2=\frac{2}{\cos^2(a\pi)}\left(1-\frac{1}{4a}\right)$
    & $b^2 = 4 \left( 1-\frac{1}{4a}\right)$\\
$\omega=\frac{1}{6}$
    & $b^2=\frac{3}{2\cos^2(a\pi)}\left(1-\frac{1}{6a}\right)$
    & $b^2 = \frac{\left(1-\frac{1}{6a}\right)}{\left[2+\left(1+2
       \cos\left(2\pi a \left(1-\frac{b^2}{4}
       \right)\right)\right)\right]}$\\[0.2in]
\hline
\end{tabular*}
\caption{\label{tab:exactt} Some exact expressions for indicator curves
 for even-period
  orbits.}
\end{center}
\end{table*}
Shinohara and Aizawa\cite{shinohara98} used the numerical observation
that at the point of hyperbolic collision (Fig.~\ref{fig:evenstd}b)
 for even-period orbits (but not for
the odd-period case), two of the indicator points {\it belong} to the
hyperbolic periodic orbit. This implies that (for period $2n$)
\begin{equation}
M^{2n}\vect{z}_j^{(0,1)} =\vect{z}_j^{(0,1)}
\label{eq:ind_curve2}
\end{equation}
for either $j=0$ or $j=1$, where
$\vect{z}_j^{(0,1)}$ are given in \eqeqref{shin_pts}.
The symmetries of the map further imply that the indicator points map
onto each other after $n$ iterations, \ie
\begin{equation} \label{eq:ind_curve}
M^n\vect{z}_j^{(0,1)} = \vect{z}_j^{(1,0)}\,.
\end{equation}
By solving these  equations for the two unknowns $(a,b)$, we can
obtain exact expressions for the bifurcation thresholds. Some of the
resulting curves for low-period orbits are given in Table \ref{tab:exactt}.
They are related to the ones given in \rcite{shinohara98} by a simple
transformation.

We will later see that when more than two chains of periodic orbits
exist, the collision threshold for some of them, but not for the
others, is given by the above criterion. Hence
we will find it useful to introduce the notion of {\it indicator curves} $b =
\Psi_{m/n}(a)$ defined by (\ref{eq:ind_curve2}) (or (\ref{eq:ind_curve})) .
When only two chains of $m/n$-orbits exist, either $\Phi_{m/n,1}$ or
 $\Phi_{m/n,3}$ coincides with  $\Psi_{m/n}$. Hence we label indicator
curves by symmetry lines, e.g., $\Psi_{m/n,1}$. (We will denote these curves
by ``ic1'' and ``ic3'' in the figures for clarity.)

\begin{remark}
Shinohara and Aizawa also proposed a numerical criterion to find the
odd-period reconnection threshold. They discovered numerically that
at the threshold successive iterates of indicator points approach
the same hyperbolic periodic point of the reconnecting chains. For
details see \rcite{shinohara98}.
\end{remark}

\subsection{Meandering orbits}
\label{ssec:nt-meanders}

Another characteristic of nontwist maps is the occurrence of {\em
meandering orbits}, which are readily observed in the standard
reconnection-collision scenario for odd-period orbits, but not in the
one for even-period orbits. (For non-standard even-period
reconnection-collision scenarios in which meanders occur,
see~\secref{evenreco}.)  As seen in Fig.~\ref{fig:oddstd}c (and
also in \figref{samplexy}), in the region surrounding $\gamma_S$,
confined by two dimerized chains, new periodic orbits and non-KAM tori
appear, \ie orbits and tori that did not exist at zero perturbation ($b=0$).
These tori are not graphs over the $x$-axis and have been called
\emph{meanders} or \emph{meandering curves}.\cite{simo98,
vanderweele88} Such invariant tori can occur only in nontwist maps because
any invariant torus for a {\it twist} map must be a graph over $x$.

It is observed numerically that in the {\it meandering region}
the winding numbers of the meandering orbits are less than the winding
 number of the reconnecting periodic orbits. Conversely, however, a ``valley''
 in the winding number profile does not imply the existence of meanders as
seen, \eg in Fig.~\ref{fig:oddstd}e. The existence of the local
minimum and the valley leads to four or more chains of periodic orbits
for certain winding numbers. We will discuss in \secref{evenreco} and
\secref{oddreco} the reconnection and bifurcation phenomena in this
scenario, which we call the \emph{non-standard scenario}.

\subsection{Break-up diagram}
\label{ssec:nt-breakup}

Since the shearless curve $\gamma_S$, whenever it exists, poses a
barrier to transport between the two twist regions, studying its
break-up is of considerable practical interest. Highly detailed
studies of break-up of shearless curves were conducted for a few
winding numbers using Greene's residue criterion.\cite{apte03, apte04,
del_castillo96} Though the method is very precise, it is not suitable
for an exploration of all parameter space.  Shinohara and
Aizawa\cite{shinohara97} obtained a rough estimate for the break-up
threshold of many shearless curves by investigating for a range
of parameter values whether iterates of one
of the indicator points remain bounded.

Here, we implement the following slightly different strategy, using
the fact that if the winding number of the orbit of any point
exists, then the orbit is not chaotic -- it is either periodic or
quasiperiodic. We calculate the sequence $\omega_i = x_i/i$ for the
iterates $(x_i,y_i) = M^i(\vect{z})$ of one of the indicator points
$\vect{z}_j^{(0,1)}$. The winding number is assumed to exist if we
can find some $N$ such that $|\omega_i-\omega_{i+1}|<\epsilon,\
\sup_{n < N}\ \omega_n > \omega_i $, and $\inf_{n < N}\ \omega_n <
\omega_i$ for $N<i<N+M$. We use $\vect{z} = (a/2+1/4,0),\ \epsilon =
10^{-7},\ M = 10^5$, and the maximum $N$ used is $2.9\times 10^6$.
If the winding number sequence displays larger fluctuations, we
assume the orbit to be chaotic, \ie the torus to be destroyed. This
criterion is also only approximate, since the value of the winding
number might converge for the number of iterations used, but further
iterations would reveal fluctuations, or vice versa.

The method of \rcite{shinohara97} and ours deliver similar results.
However, it seems that the computation of the winding number provides
better means of monitoring
and controlling its accuracy (aside from giving us the winding number
of the shearless curve as a useful side-product).

The boundary of the resulting {\it break-up diagram} displays a
fractal-like structure (see, e.g., \figref{shin_plot}). The analysis
of \rcites{apte03, del_castillo96, shinohara97} indicates that the
highest peaks correspond to the break-up of shearless invariant tori
with noble winding numbers (see \rcites{apte03, apte04} for
discussion). We will comment in \secref{discussion} on the relation
between the break-up diagram, reconnection, and the indicator and
bifurcation curves.

\section{Non-standard reconnection and collision scenarios}
\label{sec:nonstd}

As noted in \ssecref{nt-meanders}, the winding number profile can
have a valley and a local minimum (denoted by
$\omega_\text{min}$) between two local maxima (denoted by
 $\omega_\text{max} \leq \omega'_\text{max}$).
 (In many cases it is observed that $\omega_\text{max}
 = \omega'_\text{max}$, so we will not distinguish
between the two maxima.) Thus there are {\it four} orbits for each winding
number in the range $ \omega_\text{min} < \omega <
\omega_\text{max}$. The maxima and
the minimum are seen to decrease when the perturbation, $b$, is increased. This
gives rise to the following scenarios for reconnection and collision
of $m/n$-periodic orbits:
\begin{itemize}
\item For $\omega_\text{min} > m/n$, there are two chains of
 $m/n$-periodic  orbits.
\item When $\omega_\text{min}$ reaches the rational value $m/n$, periodic
  orbits of winding number $m/n$ are born and subsequently reconnect, in
 addition to already
  existing orbits. We call the new orbits \emph{inner} orbits while
  the already existing ones will be called \emph{outer} orbits.
\item When $\omega_\text{max}$ reaches $m/n$, the inner orbits reconnect and
  collide with the outer orbits.
\item For $\omega_\text{max} < m/n$, no $m/n$-periodic orbits exist.
\end{itemize}
The reconnections and collisions that occur in the above scenario are
locally the same as those seen in \ssecref{nt-coll}. But because of
the presence of four chains of periodic orbits, the global topology is
considerably more complicated as will be seen below.
\begin{remark}
  A valley in the winding number profile shows up after
  reconnection, together with meandering orbits, and also persists after the
  collision of the reconnecting orbits (Fig.~\ref{fig:oddstd}e).
  The scenario described in the
  above paragraph typically occurs for parameter values slightly above
  (in $b$ vs.\ $a$ parameter space plots) the
  bifurcation curves of odd-period orbits.
\end{remark}
\begin{remark}
  When a valley in the winding number profile exists, the winding number
  of the shearless curve is the local minimum
 $\omega_\text{min}$, if
$\omega_\text{min}$ is irrational.
\end{remark}
\begin{remark}
  When $\omega_\text{min} = m/n$, the reconnection process
 of the inner (meandering) orbits involves orbits, called {\em second-order
  meanders},\cite{simo98} that meander around these inner orbits. The
  winding number profile shows a ``hill'' at the bottom of the
  valley. Such a process might give rise to arbitrarily
  higher order meanders.\cite{petrisor01,simo98,wurm04}
\end{remark}

\subsection{Non-standard scenarios for even-period orbits}
\label{sec:evenreco}

A first hint that for the non-generic SNM the even-period reconnection
scenario can be more complicated is found in \rcite{shinohara98}.  In
some region of $(a,b)$-parameter space, phase space portraits show the
appearance of meanders and additional periodic orbits.
\begin{figure} \begin{center}
\includegraphics[width=3in]{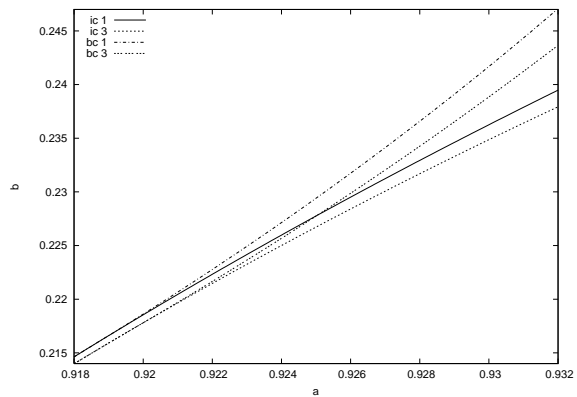}
\caption{Indicator curves (ic1 and ic3) and bifurcation curves (bc1
  and bc3) for 7/8 periodic orbits.}
\label{fig:indbifcurves} \end{center} \end{figure}
For most regions of $(a,b)$-space the indicator and
bifurcation curves are seen to coincide. But in regions of parameter
space where the winding number profile has a valley, bifurcation
curves and indicator curves separate (and can even cross each
other) leading to various reconnection scenarios and the appearance of
meanders.

This is seen most clearly near $a=1$, which is the bifurcation curve
for the $1/1$-periodic orbits, but, as mentioned above, occurs above
{\it every} odd-period bifurcation curve. We discuss in detail the example of
$7/8$-periodic orbits, whose indicator and bifurcation curves are shown
in \figref{indbifcurves} for the parameter space region of interest.
\begin{figure*}[t] \begin{center}
\includegraphics[width=0.7\textwidth]{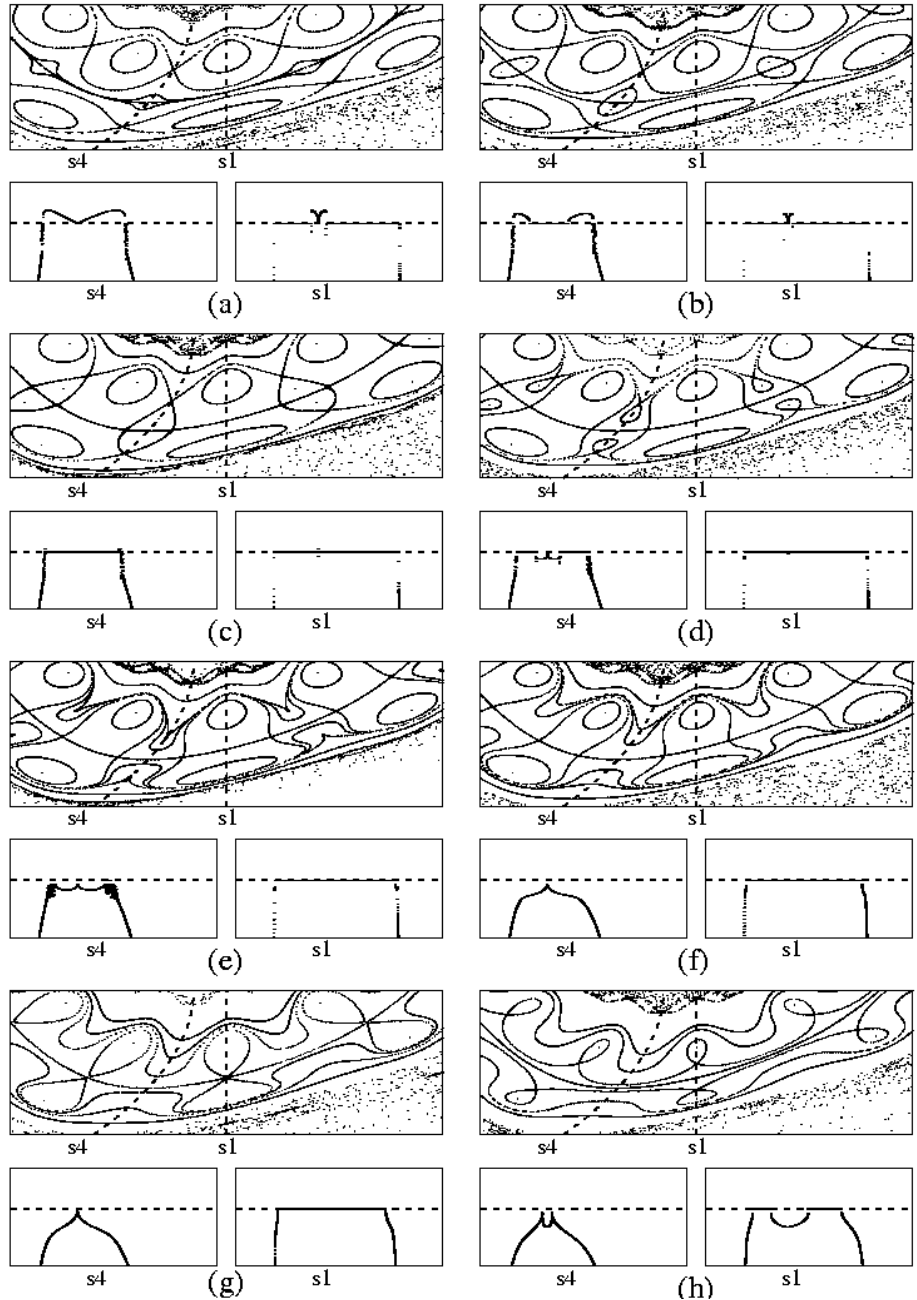}
\caption{7/8-orbit reconnection at $a=0.923$ (discussion in text).
 The upper plot shows a $y$ vs.\ $x$ phase space
plot ($x\in [-0.25,0.25] , y\in [-0.5,0]$) and the lower one the
 corresponding winding number profiles along the two indicated symmetry lines ($\omega$ along $s_i$ vs.\ $y$, $\omega\in [7/8-0.0007,7/8+0.0005]$). The
plots correspond to the $b$-values:  a) 0.22325262 $(\approx \Psi_3)$,
 b) 0.22335, c) 0.22346060, d) 0.22355, e) 0.22364169 $(\Phi_3)$, f) 0.22385,
 g) 0.22416343 $(\Psi_1)$, h) 0.22465.} 
\label{fig:evenbeforecross}
\end{center}
\end{figure*}
For approximately $a < 0.92$, the curves coincide, resulting in the
 standard scenario (\ssecref{nt-coll}). The lower curve, corresponding
to bc3 and ic3, is the
threshold for collision (i.e., reconnection) of hyperbolic orbits, while
the upper curve, corresponding to bc1 and ic1, is for the simultaneous collision
 (annihilation) of elliptic and non-symmetric
hyperbolic orbits. For $a > 0.92$ the curves separate, and around $a =
0.925$ the bifurcation curve bc3, crosses the indicator curve
ic1.

Recall that $b=\Phi_{7/8,i}(a)$ is the bifurcation curve
for orbits on the symmetry line $s_i$, and $b=\Psi_{7/8,i}(a)$ is the
indicator curve. In the
following we drop the subscript $7/8$ and the dependence on $a$. Also,
the symmetry of the SNM implies that $\Phi_1 = \Phi_2$ and $\Phi_4 =
\Phi_3$.

The separation gives rise to the following two sequences:

For $0.92 < a < 0.925$, the above curves are in the following order:
$\Psi_3 < \Phi_3 < \Psi_1 < \Phi_1 $.
\begin{enumerate}
\item\label{even1_1} Whereas for small perturbations ($b < \Psi_3$),
  only the up and down (outer) $7/8$-chains exist
  ($\omega_\text{min}>7/8$), $\omega_\text{min}$ reaches $7/8$ and
  inner orbits are born at $b = \Psi_3$
  (Fig.~\ref{fig:evenbeforecross}a).  Initially, this inner chain has
  the dipole topology, with elliptic orbits on $s_3$ and $s_4$, and
  hyperbolic orbits not on any symmetry
 line. (Fig.~\ref{fig:evenbeforecross}b).
\item\label{even1_2} Somewhere in the range $ \Psi_3 < b <
  \Phi_3 $, a reconnection between these off-symmetry line hyperbolic
  orbits with the outer hyperbolic orbits on $s_3$ and $s_4$ occurs
  (Fig.~\ref{fig:evenbeforecross}c). At the point of reconnection,
  $\omega_\text{max}$ is $7/8$ and continues to be so through step
  \ref{even1_5}. After this reconnection, meanders are born, the inner
  and outer chains display a nested topology, and $\omega_\text{min} <
  7/8$ (Fig.~\ref{fig:evenbeforecross}d).
\item\label{even1_3} The inner elliptic orbits collide with the outer
  hyperbolic orbits on the $s_3$ and $s_4$ symmetry line at $b =
  \Phi_3$ (Fig.~\ref{fig:evenbeforecross}e), leaving only the
  off-symmetry line inner hyperbolic orbits and the outer elliptic
  orbits on the symmetry lines $s_1$ and $s_2$
  (Fig.~\ref{fig:evenbeforecross}f).
\item\label{even1_4} At $b = \Psi_1$, the inner hyperbolic orbits
  collide (Fig.~\ref{fig:evenbeforecross}g) and
  move onto the symmetry lines $s_1$ and $s_2$
  (Fig.~\ref{fig:evenbeforecross}h).
\item\label{even1_5} At $ b = \Phi_1 $, there is an
  hyperbolic-elliptic collision on $s_1$ and $s_2$, after which
  $7/8$-periodic orbits cease to exist and $\omega_\text{max} < 7/8$.
\end{enumerate}
For $a > 0.925$, the order of the curves is changed to $\Psi_3 <
\Psi_1 < \Phi_3 < \Phi_1 $.  Most steps of the above sequence
remain the same (with $\Psi_1$ replacing $\Phi_3$ in the second step),
except for the following:
\begin{enumerate}
\setcounter{enumi}{2}
\item\label{even2_3} At $b = \Psi_1$, the inner hyperbolic orbits
  collide and move onto the symmetry lines $s_1$ and $s_2$, while the
  hyperbolic elliptic orbits on $s_3$ and $s_4$ merely continue to
  approach each other.
\item\label{even2_4} The inner elliptic orbits collide with the outer
  hyperbolic orbits on $s_3$ and $s_4$ at $b = \Phi_3$, leaving only
  the hyperbolic and elliptic orbits on $s_1$ and $s_2$.
\end{enumerate}
The difference between the two scenarios is the order in which the collision
along $s_3$ and $s_4$ and the move of the hyperbolic points onto
 $s_1$ and $s_2$ occur. Both occur simultaneously for the $(a,b)$-value
 at which $\Phi_3$ and $\Psi_1$ intersect.

As in the standard scenario of odd-period orbits, meanders appear in
this case when the maximum of the winding number profile is rational.
But now there are {\it two} meandering regions, and none of the meanders are
are $\mathcal{G}$-invariant (Fig.~\ref{fig:evenbeforecross}e, along
$s_4$).

Note that the indicator curves, which are obtained using the
symmetries of the map, track collisions occurring 
along the $\mathcal{G}$-invariant curve while the bifurcation curves 
track the collisions occurring at maxima in the winding number profile,
 which may or may not be shearless.

This motivates an extension of the definition of bifurcation curves.
Recall that in the original definition the bifurcation curve
determines the {\it global} boundary in parameter space between
existence and non-existence of periodic orbits of  the corresponding
winding number. These curves correspond to maxima in the winding
number plot, and will be denoted by $\Phi^\text{out}_{m/n,i}$ from now on.

In the region of parameter space where four (or more) orbits of a particular
winding number exist, a {\it new} kind of bifurcation curve can be defined
that corresponds to parameter values at which the inner orbits are
born, denoted by $\Phi^\text{in}$. In this case, the winding number
profile has a {\it local} minimum (or {\it local} maximum for
higher order curves). For even-period orbits, as expected and verified
numerically,  $\Psi_i = \Phi_i^\text{in}$ .
\begin{figure*}[t] \begin{center}
\includegraphics[width=0.55\textwidth]{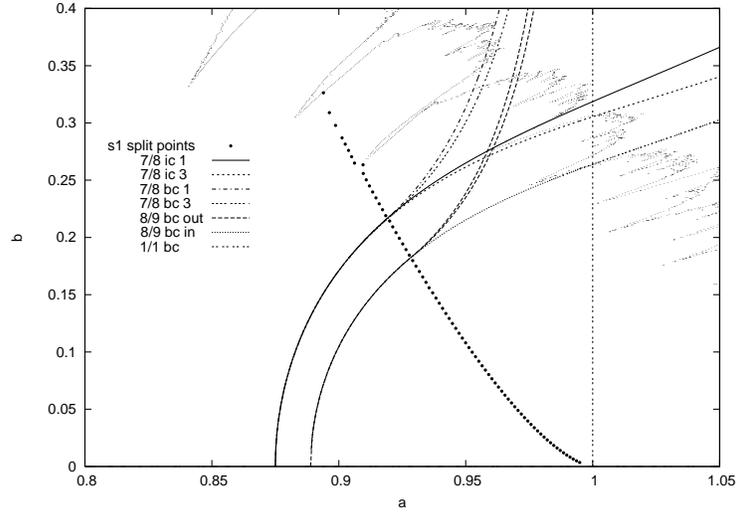}
\caption{Boundary of break-up diagram (dots), indicator curves (ic)
  and bifurcation curves (bc) for two different winding
  numbers. Points marked ($\bullet$) correspond to points where
 $\Phi^\text{out}_1$ and $\Phi^\text{in}_1$ of close-by periodic
orbits separate.}
\label{fig:shinplot2} \end{center} \end{figure*}
\begin{figure} \begin{center}
\includegraphics[width=3in]{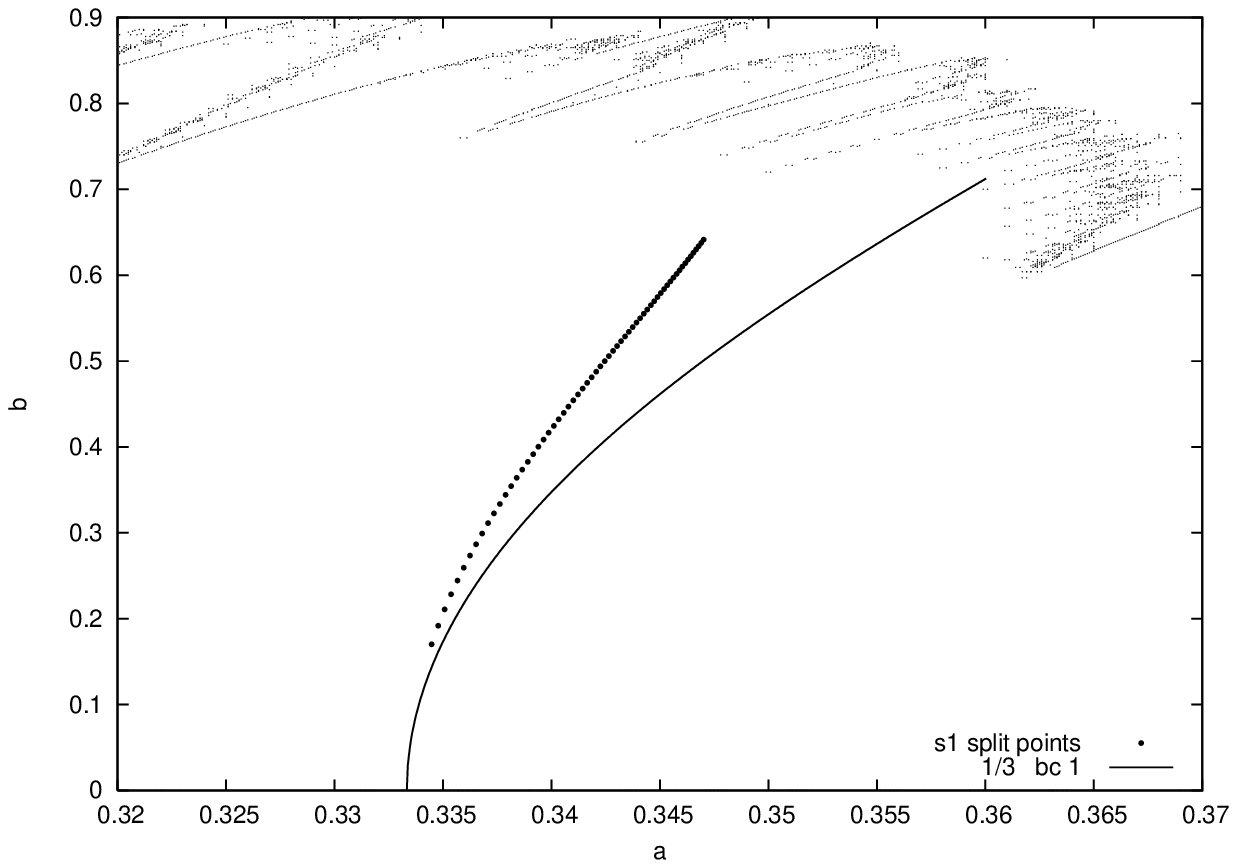}\\
\includegraphics[width=3in]{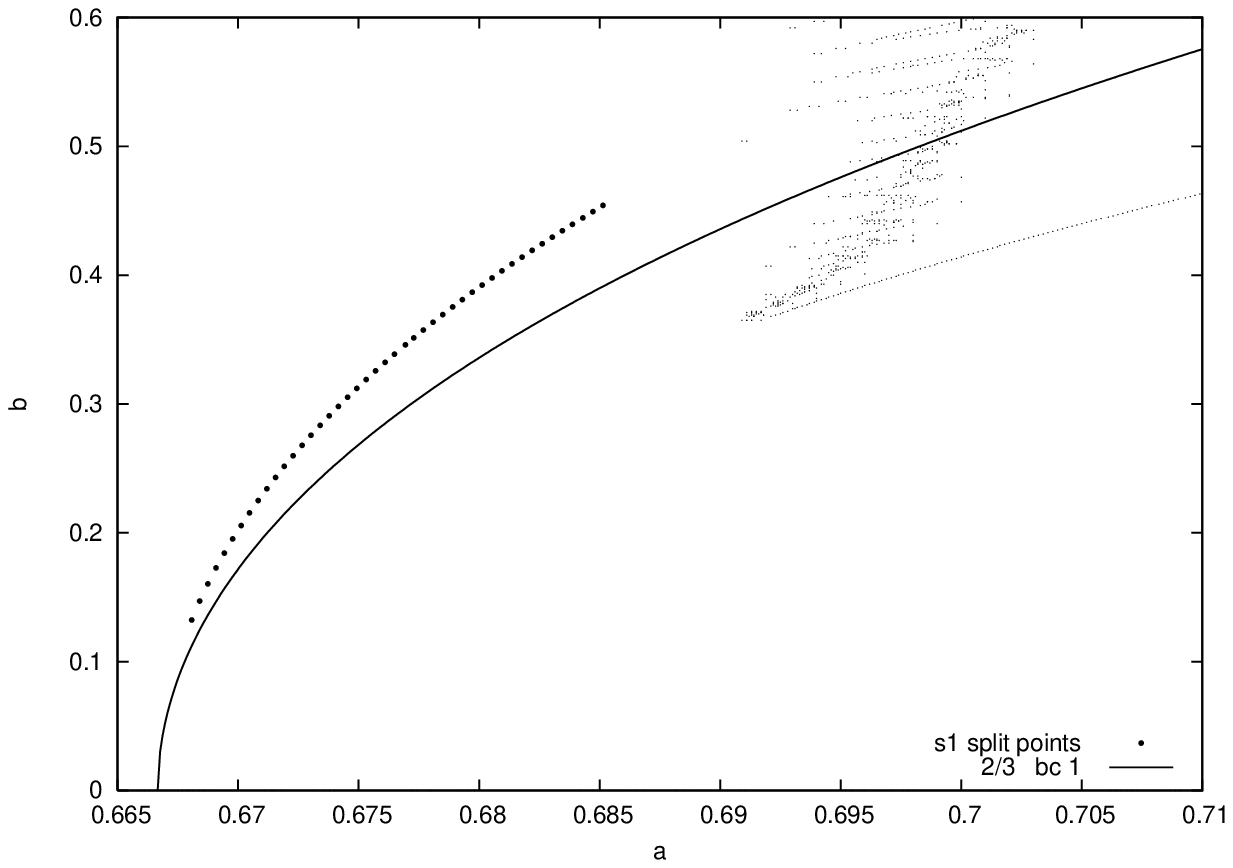}
\caption{Boundary of break-up diagram (dots) and bc1 for
$1/3$-orbits (top) and $2/3$-orbits (bottom). Points marked
($\bullet$) correspond to points where $\Phi^\text{out}_1$ and $\Phi^\text{in}_1$ of close-by periodic orbits separate.}
\label{fig:shinplot4} \end{center} \end{figure}
To estimate the region of parameter space in which there are more than
two chains of periodic orbits, we plot the points of separation of
indicator and bifurcation curves for several different (odd and even) winding
numbers. The results are shown in \figref{shinplot2}.  Within
numerical uncertainty, close to the $1/1$-bifurcation curve the
 points seem to lie on a curve. Similar
``curves'' have been numerically observed for a few other winding
numbers, e.g., \figref{shinplot4}, where we mark the points at which the
indicator curves split from the bifurcation curves (along $s_1$) for
several inner orbits, in the region of parameter space
close to the $1/3$-orbit (left) and $2/3$-orbit (right) bifurcation curve (also shown).
\begin{remark}
  In \rcite{petrisor03} Petrisor {\it et al.\ } conjectured that in
 generic nontwist maps the dipole formation scenario for even-period
 orbits does not occur. Instead, their numerical studies showed that
 one of the elliptic chains bifurcates, creating and subsequently
 destroying a saddle-center pair, which has the effect of aligning the
 elliptic points of one chain with the hyperbolic points of the other
 one. Reconnection occurs then according to the standard scenario for
 {\it odd}-period reconnection.
\end{remark}

\subsection{Non-standard reconnection for odd-period orbits}
\label{sec:oddreco}

\begin{figure} \begin{center}
\includegraphics[width=3in]{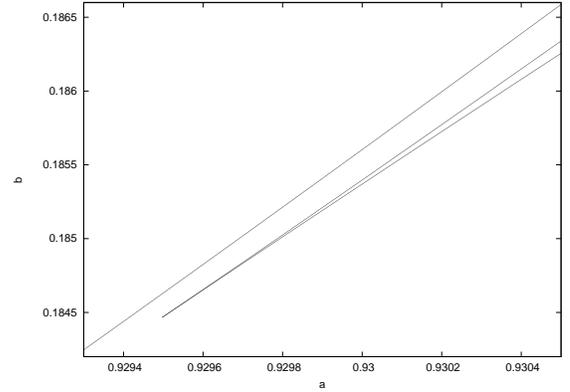}
\caption{Magnification of split region in parameter space for 8/9-bifurcation
curves. The curves appear in the order $\Phi^\text{in} < \Phi^\text{out,up}_1
< \Phi^\text{out,down}_1$. (discussion in text)}
\label{fig:odd_bifs} \end{center} \end{figure}
Since there is no analog of the criterion \eqeqref{ind_curve} for the
odd-period orbits, we cannot define indicator curves for these
orbits. But we note that the indicator curves for even-period orbits are the bifurcation curves for the inner orbits, $\Psi_i = \Phi_i^\text{in}$. Thus, we have implemented a
 version of the numerical method for
finding bifurcation curves\cite{apte03,del_castillo96} for {\it both} the inner and
outer orbits in the parameter space region where more than two chains
exist. \figref{shinplot2} shows the bifurcation curves for the outer
orbits, $b = \Phi_{8/9,i}^\text{out}(a)$, and inner orbits, $b =
\Phi_{8/9,i}^\text{in}(a)$, of winding number $8/9$. (Recall that in
the odd-period standard scenario the bifurcation curves, $\Phi^\text{out}_i$,
are the same for all four symmetry lines.) For the remainder of this section,
 we will drop the dependence on $a$ and the subscript $8/9$ for brevity.

A magnification of $(a,b)$-space around the separation point
(\figref{odd_bifs}) reveals the birth of a {\it new} hyperbolic-elliptic
pair of periodic orbits along each of the symmetry lines at $b=\Phi^\text{in}$
for $a > 0.9295$ (lowest curve in \figref{odd_bifs}). This pair of
orbits moves apart and eventually collides/annihilates with the outer, up
and down, orbits on the symmetry lines. These collisions do not occur
simultaneously in parameter space (in contrast to the even case),
 which explains the existence
of {\it two} outer bifurcation curves along each symmetry line, one
for the outer up orbit and one for the outer down orbit, denoted by
$\Phi^\text{out,up}_i$ and $\Phi^\text{out,down}_i$, respectively.
Because of the symmetry of the SNM, $\Phi^\text{out,up}_\text{1 and 2}
= \Phi^\text{out,down}_\text{3 and 4}$ and $\Phi^\text{out,down}_\text{1 and 2}
= \Phi^\text{out,up}_\text{3 and 4}$.
\begin{figure*}[t] \begin{center}
\includegraphics[width=0.7\textwidth]{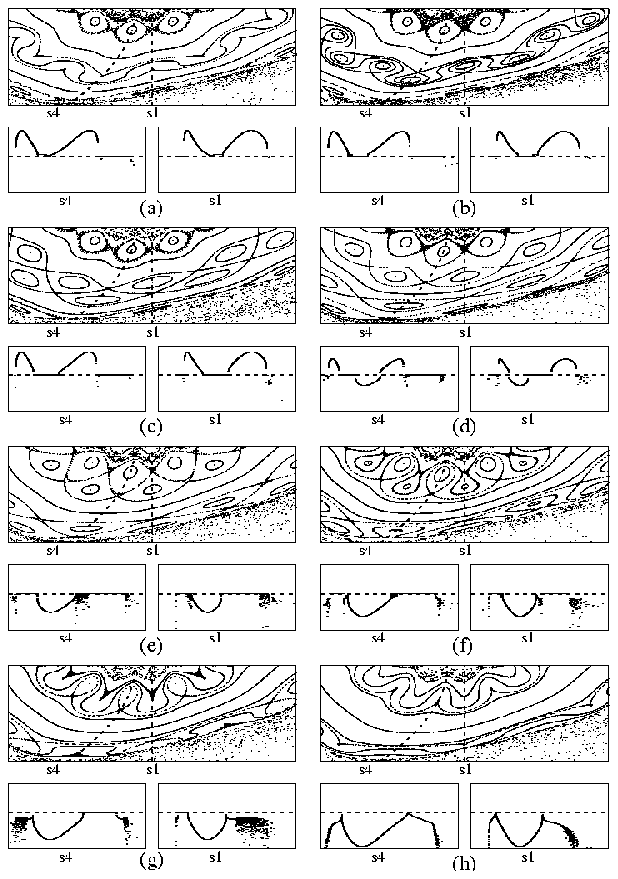}
\caption{8/9-orbit reconnection at $a=0.94$ (discussion in text).
The upper plot shows a $y$ vs.\ $x$ phase space
plot ($x\in [-0.25,0.25] , y\in [-0.5,0]$) and the lower one the
 corresponding winding number profiles along the two indicated symmetry lines ($\omega$ along $s_i$ vs.\ $y$, $\omega\in [8/9-0.0020,8/9+0.0016]$). The
plots correspond to the $b$-values:  a) 0.2015781 $(\approx \Phi^\text{in})$,
 b) 0.2018, c) 0.202155, d) 0.2036, e) 0.20515, f) 0.2059, 
g) 0.2067917 $(\approx \Phi_1^\text{out,up})$, h) 0.2081084 $(\approx
\Phi_1^\text{out,down})$.}
\label{fig:oddbeforecross}
\end{center}
\end{figure*}
\begin{figure} \begin{center}
\includegraphics[width=2.5in]{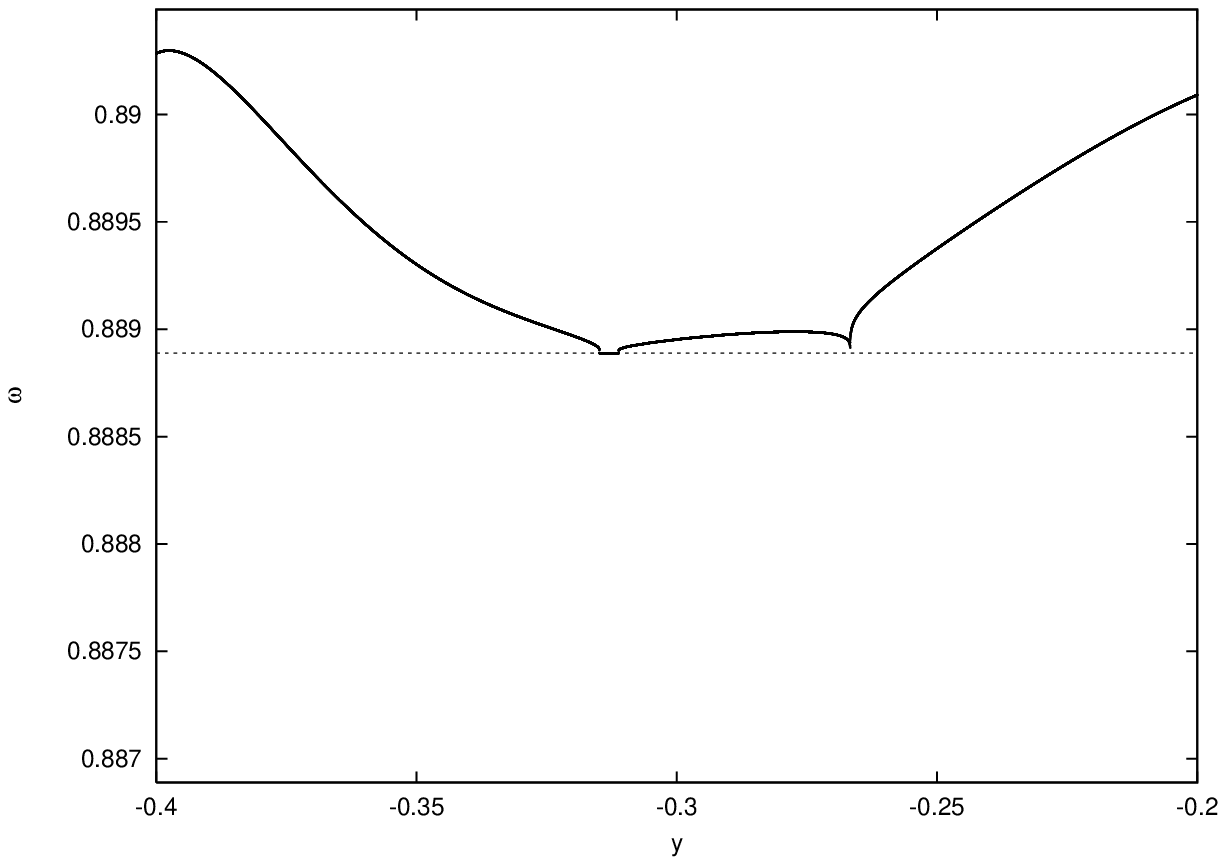}\\
\includegraphics[width=2.5in]{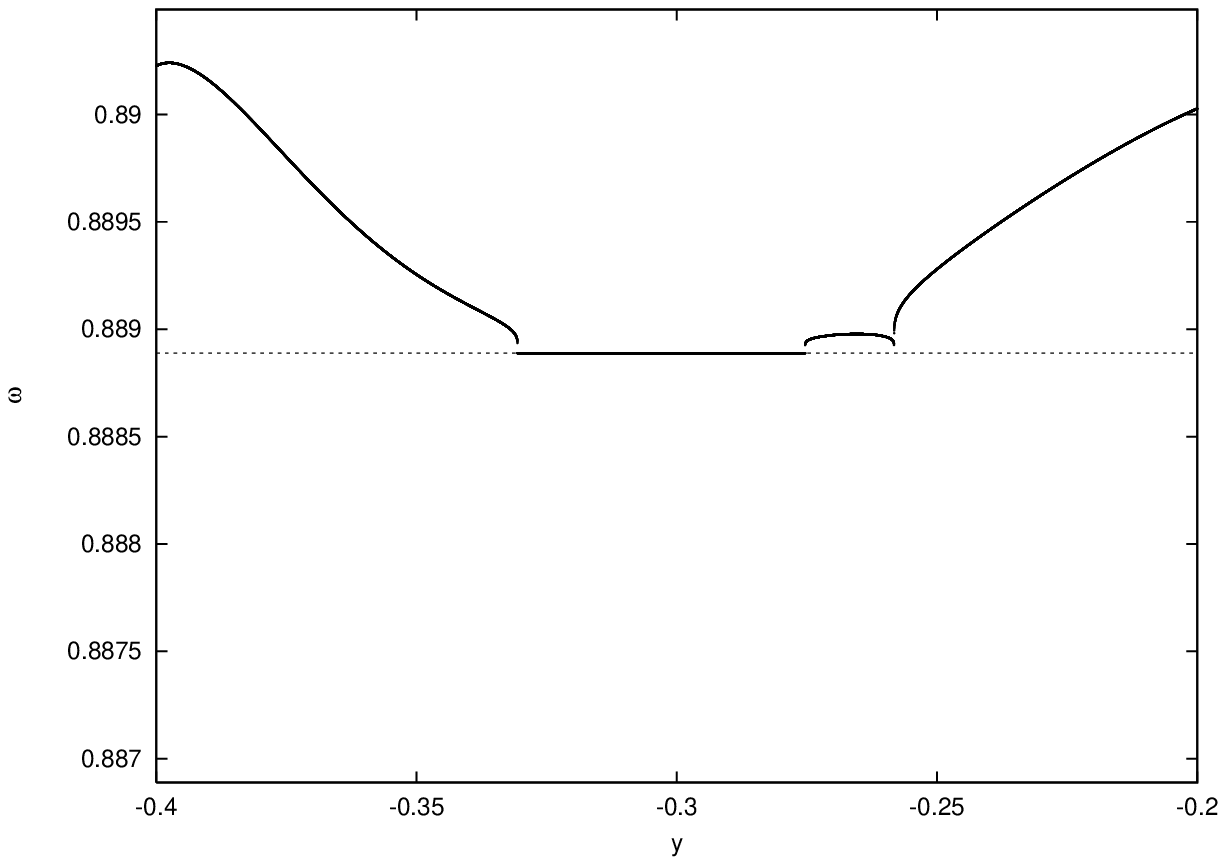}
\caption{Magnification of the winding number profiles along $s_1$ of
Figs.~\ref{fig:oddbeforecross}a (top)  and~\ref{fig:oddbeforecross}b
(bottom)..
\label{fig:zoomed}}
\end{center}
\end{figure}
To illustrate this non-standard reconnection-collision sequence, we
discuss in some detail the case of $8/9$-periodic orbits along $s_1$
at $a=0.94$ (Fig.~\ref{fig:oddbeforecross}):
\begin{enumerate}
\item Whereas for small perturbations ($b < \Phi^\text{in}$), only the
  up and down (outer) orbits exist and $\omega_\text{min} > 8/9$,
  inner chains of orbits are born (Fig.~\ref{fig:oddbeforecross}a) at
  $b = \Phi^\text{in}$, and subsequently form a nested topology with
  meandering orbits (Fig.~\ref{fig:oddbeforecross}b).  As seen in the
  corresponding winding number profile along $s_1$, the bottom of the
  valley exhibits a hill whose edges have the value
  $\omega_\text{min}=8/9$.  The left edge, corresponding to the new
  periodic orbit pair, broadens as its hyperbolic and elliptic orbits
  start to move apart (Fig.~\ref{fig:zoomed}), whereas the right one,
  corresponding to the heteroclinic connection between new pairs on
  other symmetry lines, continues to touch $\omega_\text{min}=8/9$ at
  only one point.  The winding number of the shearless curve,
  corresponding to the {\it local} maximum (top of hill), and those of
  the meandering orbits are greater than $8/9$.
\item Somewhere in the range $ \Phi^\text{in} < b <
  \Phi^\text{out,up}_1$, the nested orbits reconnect
  (Fig.~\ref{fig:oddbeforecross}c), resulting in four regular
  Poincar\'e-Birkhoff chains (Fig.~\ref{fig:oddbeforecross}d), two on
  each side of the central shearless curve.
\item At a higher $b$-value in the same range, the two up chains and
  two down chains reconnect (Fig.~\ref{fig:oddbeforecross}e),
  resulting in two regions of nested orbits with meandering tori
  (Fig.~\ref{fig:oddbeforecross}f). Again, the winding numbers of the
  meanders is less than that of the reconnecting orbits ($8/9$).
  Each of these regions contains a shearless, but not $\mathcal{G}$-invariant,
   meander.
\item At $b = \Phi^\text{out,up}_1$, the outer up orbits undergo a
  hyperbolic-elliptic collision (Fig.~\ref{fig:oddbeforecross}g).
\item At $b = \Phi^\text{out,down}_1$, the outer down orbits undergo a
  hyperbolic-elliptic collision (Fig.~\ref{fig:oddbeforecross}h),
  after which no further $8/9$-periodic orbits exist.
\end{enumerate}

\section{Discussion}
\label{sec:discussion}

\begin{figure*}[t] \begin{center}
\includegraphics[width=0.7\textwidth]{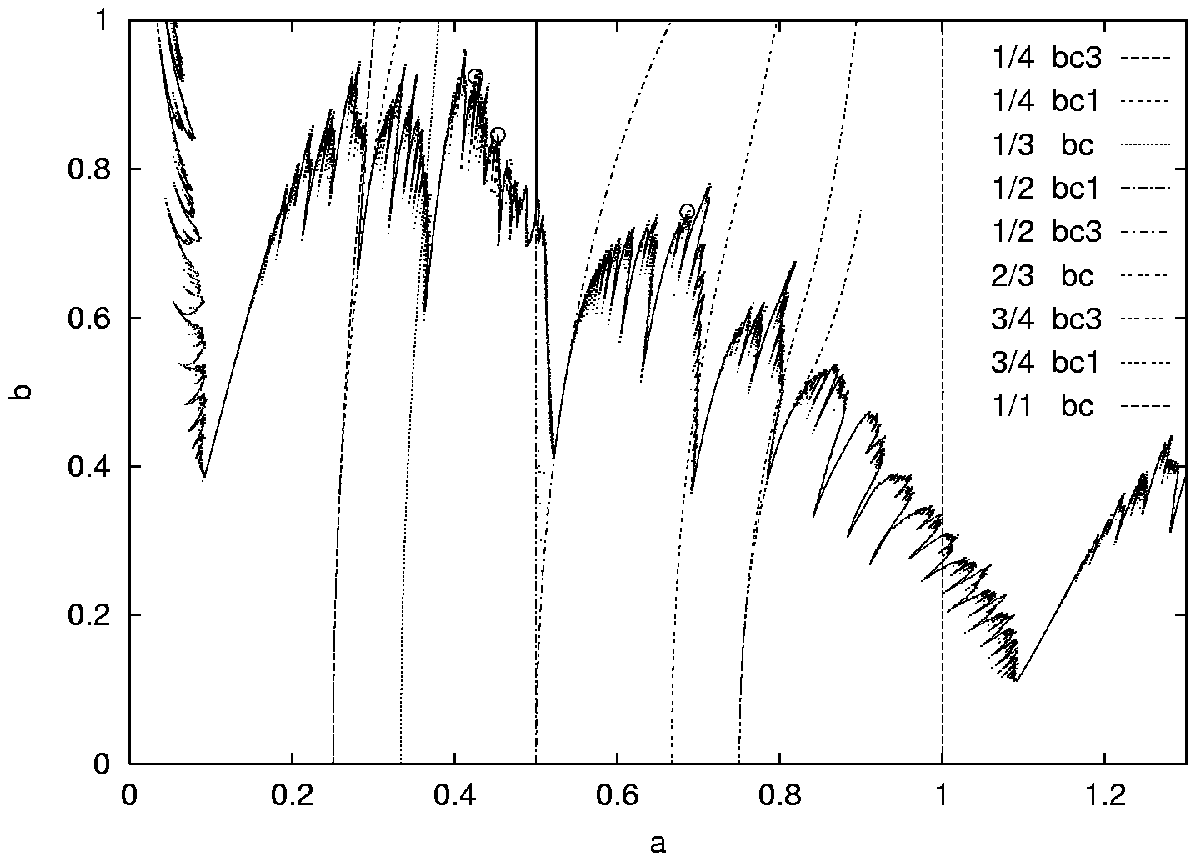}
\caption{Parameter space showing the points for which shearless
  invariant tori exist. Also shown are some bifurcation and
  indicator curves and the critical
  points (marked by $\circ$) found using Greene's residue criterion.}
\label{fig:shin_plot} \end{center}
\end{figure*}

To understand the effect of the different reconnection-collision
scenarios on the global transport properties of the map, we compare
the computed bifurcation and indicator curves with the break-up
diagram, obtained using the procedure of \ssecref{nt-breakup}. The
results are presented in  \figref{shin_plot}. We see that these
curves serve as a scaffolding for the diagram, an observation also made in
\rcites{shinohara98,wurm04}.
\begin{figure*}[t] \begin{center}
\includegraphics[width=0.7\textwidth]{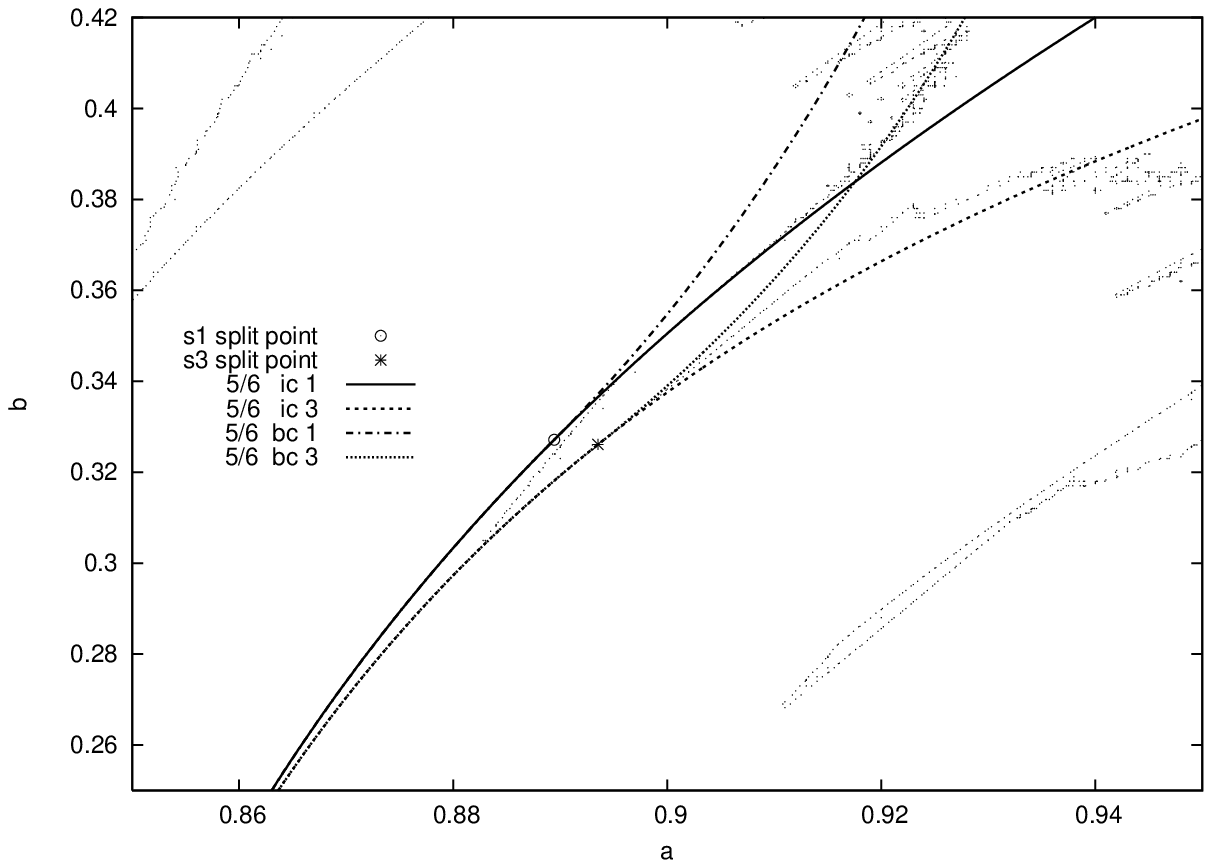}
\caption{Section of parameter space around the $5/6$-orbit indicator and bifurcation curves.}
\label{fig:shin_plot3} \end{center}
\end{figure*}
We will illustrate this interplay
between the different reconnection scenarios and transition to
global chaos by looking at the
 magnification of the break-up diagram around the indicator and bifurcation
curves of the $5/6$-orbits (\figref{shin_plot3}).
\begin{itemize}
\item For approximately $a<0.883$, indicator and bifurcation curves
along the different symmetry lines coincide, and lie within the
non-chaotic region of the break-up diagram. Fixing $a$ and increasing
the perturbation $b$ through the reconnection process does not lead to
global chaos. This is because of the presence of invariant tori
``above'' and ``below'' the reconnecting chains of orbits. (For the
remaining discussion, we will refer to these as "twist" tori.)
\item For approximately $0.883 < a < 0.893$, the lower pair of curves
follows the boundary of the break-up diagram.  Increasing the
perturbation through the boundary corresponds to a reconnection that
results in global chaos. The reason for this can be understood as
follows: For parameter values slightly below the lower pair of curves,
no twist invariant tori exist above or below the shearless region. Only the tori in the region between the two chains of $5/6$-orbits inhibit transport.
When the perturbation $b$
reaches the lower pair of curves, the hyperbolic $5/6$-orbits
collide and the tori between the two chains are destroyed. Since the
colliding orbits are hyperbolic, their separatrices
form a heteroclinic tangle that connects the up and down chaotic
regions resulting in global chaos, consistent with the observation that the lower pair of curves form a boundary of the break-up diagram.
\item For approximately $0.893 < a < 0.91$, we see the first of the
two non-standard scenarios (Fig.~\ref{fig:evenbeforecross}) discussed
in \secref{evenreco}. Here again, no twist tori exist for parameter
values close to the lowest curve $\Psi_3$. The inner orbits, which are born
at $b=\Psi_3$, move apart and reconnect with the outer
hyperbolic orbits for some $\Psi_3<b<\Phi_3$. At this point, global chaos ensues
because the up and down chaotic regions are joined through the
reconnecting invariant manifolds. Thus the lower boundary of the break-up
diagram lies between $\Psi_3$ and $\Phi_3$.\\
Also, for $b>\Psi_1$, the off-symmetry line hyperbolic orbits move
onto the $s_1$ symmetry line and their invariant manifolds are no
longer connected (``de-connection'').  Meandering orbits are created
as these hyperbolic orbits move apart along $s_1$. These meandering
orbits (Fig.~\ref{fig:evenbeforecross}h) inhibit global chaos. Thus,
$b=\Psi_1$ (seen as the threshold for ``de-connection'') forms the
upper boundary of the break-up diagram.
\item It was noted in Ref.~\cite{wurm04} that if the up and down twist
regions of the map exhibit chaos when the odd-period orbits reconnect,
the reconnection leads to global chaos for the above-mentioned reason
(i.e., because of the heteroclinic tangle of the invariant manifolds).
\end{itemize}
Thus we see that, in general, reconnection (and the break-up of twist
tori) has a significant effect on the transition to global chaos. Many, if
not all, of the ``smooth'' boundaries of the break-up diagram can be
conjectured to be the result of reconnections of invariant manifolds. Indeed,
the reconnections of orbits with higher periods can give rise to finer
structures in the break-up diagram. Thus, the ``fractal'' nature of
this diagram might be related to the reconnections in 
phase space happening at smaller and smaller scales. A reliable 
criterion for determining the reconnection threshold and its numerical
implementation will be extremely useful for testing these ideas.

\section{Conclusion}
\label{sec:conclusion}

In this paper we presented a unified view of various bifurcations and
reconnections that occur in the standard nontwist map, but are not
observed in twist maps. The bifurcations are locally of three types:
collision and annihilation of hyperbolic-elliptic orbit pairs,
collision of two symmetric hyperbolic orbits resulting in two
non-symmetric hyperbolic orbits, and the simultaneous (in parameter
and phase space) collision and annihilation of two hyperbolic and two
elliptic orbits. The latter is seen only in the SNM (and closely
related maps) because of its high degree of symmetry. The two former
bifurcations are observed in all nontwist maps.

We have demonstrated that the non-generic standard nontwist map has
more types of periodic orbit reconnection scenarios than previously
known, resulting from the presence of four (or more) chains of
periodic orbits of the same winding number. Meandering tori, 
associated with the odd-period reconnection in the SNM, have been
shown to exist for certain regions of parameter space for
non-standard even-period reconnection. Petrisor and
co-workers\cite{petrisor03} have started to compile a list of possible
reconnection scenarios for nontwist maps. Our investigation
contributes new global scenarios for the case of nontwist maps with a
reversing symmetry group, including a spatial symmetry.

We also conjecture (and present heuristic observations) that the
reconnection of hyperbolic separatrices leads to global chaos if no
``twist'' tori exist in the region ``outside'' the reconnecting chains
of orbits. Determination of the break-up threshold of the last twist tori
and a precise criterion for calculating reconnection thresholds will shed
more light on the accuracy and usefulness of these observations.

\section{Acknowledgments}
This research was in part supported by U.S. Department of Energy
Contract No. DE-FG01-96ER-54346 and by an appointment of A.~Wurm to
the U.S. Department of Energy Fusion Energy Postdoctoral Research
Program administered by the Oak Ridge Institute for Science and
Education.

\appendix

\section{Basic definitions}
\label{sec:basics}

For reference, we list a few basic definitions used throughout the main text.
An {\it orbit} of an area-preserving map $M$ is a sequence of points
$\left\{\left(x_i,y_i\right)\right\}_{i=-\infty}^{\infty}$ such that
$M\left(x_i,y_i\right) = \left(x_{i+1},y_{i+1}\right)$.
The {\it winding number} $\omega$ of an orbit is defined as
the limit $\omega = \lim_{i\to\infty} (x_i/i)$, when it exists.
Here the $x$-coordinate is ``lifted'' from $\Tset$ to $\Rset$.
A {\it periodic orbit} of period $n$ is an orbit $M^n \left( x_i,
y_i\right) = \left( x_i+m, y_i\right)$, $\forall \:i$, where $m$ is an
integer. Periodic orbits have rational winding numbers $\omega=m/n$.
An {\it invariant torus} is a one-dimensional set $C$ that is invariant
under the map $C = M(C)$. Of particular importance are the invariant
tori that are homeomorphic to a circle and wind around the $x$-domain
because, in two-dimensional maps, they act as transport
barriers. Orbits belonging to such a torus generically have irrational
winding number.

\section{Symmetry properties of the SNM}
\label{sec:symm}

In this appendix, we review the symmetry properties of the SNM. A
detailed discussion of these types of symmetries in general can be found in
\rcite{lamb98} and, in the context of nontwist systems, in \rcite{petrisor01}.

The standard nontwist map $M$ is time-reversal symmetric with respect
to two involutions $I_0$ and $I_1$, i.e.,
$M^{-1} = I_i^{-1}\ M\ I_i$, $I_i^2 = {\rm Id}$,
for $i=0,1$. It then follows that the SNM can be decomposed as $M =
I_1\circ I_0$. In addition, $M$ is also symmetric under an involution $S$
which commutes with both $I_i$, i.e.,
$M = S^{-1}\ M\ S$, $S^2 = {\rm Id}$, and $S\ I_i = I_i\ S$.
The involutions have the following form:
$I_0(x,y) = (-x          , y-b\sin(2\pi x))$,
$I_1(x,y) = (-x+a(1-y^2) , y)$, and
$S(x,y)   = (x+ 1/2      , -y)$.

Since the $I_i$ are orientation reversing, their fixed point sets,
$\Gamma_i = \left\{ (x,y)| I_i (x,y) = (x,y)\right\}$, are
one-dimensional sets, called the \emph{symmetry lines} of the map.
For the SNM, $\Gamma_0$ consists of
$s_1=\left\{(x,y) |x=0\right\}$, 
$s_2=\left\{(x,y) |x=1/2\right\}$,
while $\Gamma_1$ consists of
$s_3=\left\{(x,y) |x=a\left(1-y^2\right)/2\right\}$,
$s_4=\left\{(x,y) |x=a\left(1-y^2\right)/2+1/2\right\}$.

It is easy to check that $S$ does not have any fixed points.
Symmetry lines are useful because the numerical search for symmetric
periodic orbits, i.e., orbits with a point belonging to one of the
symmetry lines, is a one-dimensional root finding problem and hence
considerably easier than the search for non-symmetric
orbits.\cite{greene68,devogel58} For the SNM, there are generally two
orbits along any symmetry line of any winding number, called the
\emph{up} and \emph{down} orbits.

The SNM is time-reversal symmetric with respect to maps $SI_i$ as
well. The fixed point sets of these orientation preserving maps
consist of points in phase space, called \emph{indicator points}. For
the SNM, fixed points of $SI_0$ and $SI_1$ are, respectively,
$\vect{z}_0^{(0,1)}=\left(\mp\frac{1}{4},\mp\frac{b}{2}\right)$, and
$\vect{z}_1^{(0,1)}=\left(\frac{a}{2}\mp\frac{1}{4},0\right)$.

The involutions $I_0$, $I_1$, $S$, and their products form a group
$\cal G$ called the \emph{reversing symmetry group}. There is at most
one homotopically nontrivial invariant torus that is invariant under
the action of $\cal G$\cite{petrisor01}. This torus is called the \emph{central}
or \emph{$\mathcal{G}$-invariant shearless torus} $\gamma_S$.
 The indicator points  belong to $\gamma_S$ when 
it exists.\cite{shinohara97,shinohara98,petrisor01}

\end{document}